\documentclass[aps,preprint,superscriptaddress]{revtex4}
\usepackage{graphicx}
\usepackage{amsmath}
\usepackage{dcolumn}

\def\V{\textbf{V}}
\def\v{\textbf{v}}
\def\b{\textbf{b}}

\def\B{\textbf{B}}

\def\bh{\textbf{h}}
\def\E{\textbf{E}}
\def\e{\textbf{e}}
\def\pa{\partial}

\begin{document}

\title{Numerical simulation of laminar plasma dynamos in a cylindrical von K\'arm\'an flow}

\author{I. V. Khalzov}
\affiliation{University of Wisconsin, 1150 University Avenue, Madison, Wisconsin 53706 USA}
\author{B. P. Brown}
\affiliation{University of Wisconsin, 1150 University Avenue, Madison, Wisconsin 53706 USA}
\author{F. Ebrahimi}
\affiliation{University of New Hampshire, 8 College Road, Durham, New Hampshire 03824 USA}
\author{D. D. Schnack}
\affiliation{University of Wisconsin, 1150 University Avenue, Madison, Wisconsin 53706 USA}
\author{C. B. Forest}
\affiliation{University of Wisconsin, 1150 University Avenue, Madison, Wisconsin 53706 USA}

\date{\today}

\begin{abstract}

The results of a numerical study of the magnetic dynamo effect in cylindrical von K\'arm\'an plasma flow are presented with parameters relevant to the Madison Plasma Couette Experiment (MPCX). This experiment is designed to investigate a broad class of phenomena in flowing plasmas. In a plasma, the magnetic Prandtl number $Pm$ can be of order unity (i.e.,  the fluid Reynolds number $Re$ is comparable to the magnetic Reynolds number $Rm$). This is in contrast to liquid metal experiments, where $Pm$ is small (so, $Re\gg Rm$) and the flows are always turbulent. We explore dynamo action through simulations using the extended magnetohydrodynamic (MHD)  NIMROD code for an isothermal and compressible plasma model. We also study two-fluid effects in simulations by including the Hall term in Ohm's law. We find that the counter-rotating von K\'arm\'an flow results in sustained dynamo action and the self-generation of magnetic field when the magnetic Reynolds number exceeds a critical value. For the plasma parameters of the experiment this field saturates at an amplitude corresponding to a new stable equilibrium (a laminar dynamo). We show that compressibility in the plasma results in an increase of the critical magnetic Reynolds number, while inclusion of the Hall term in Ohm's law changes the amplitude of the saturated dynamo field but not the critical value for the onset of dynamo action.     
 
\end{abstract}

\maketitle

\section{Introduction}

The dynamo phenomena, where magnetic fields are self-generated by a moving and electrically conducting fluid,
is one of the most intriguing subjects of modern magnetohydrodynamics (MHD).  Dynamos have particularly important applications in astrophysics \cite{Moffatt, Brandenburg&Subramanian_2005}. Today it is widely believed that the magnetic fields of planets and stars originate from dynamo action in their interiors \cite{Ossendrijver_2003, Charbonneau_2010, Jones_2010}. As far as we know, all astrophysical dynamos are turbulent with extremely high fluid Reynolds numbers $Re$, which makes theoretical treatments very involved and realistic simulations are currently impossible. At the same time, some key physical processes likely at work in astrophysical dynamos can be revealed by considering idealized laminar dynamos related to spatially smooth stationary velocity fields at comparatively low $Re$. In particular, the theoretical study of laminar dynamos allows one to determine the critical magnetic Reynolds number, above which dynamo excitation takes place, to find the structure and magnitude of saturated dynamo field, and to understand the influence on the dynamo of different plasma effects, such as compressibility, two-fluid effects, and anisotropic transport.  

A number of laminar MHD flows appropriate for dynamo generation have been analyzed in the literature \cite{Ponomarenko,Roberts,VKS_num,TCF_num,Dudley,Moss}.  In most of these studies only the kinematic dynamo problem has been considered, in which the magnetic induction equation is solved as an eigenvalue problem for a given velocity field (not necessarily satisfying the Navier-Stokes equation) to find the growth rate of the magnetic field. The non-linear feedback of the fields on the flows is ignored in the kinematic treatment. Among the laminar flows that lead to kinematic dynamos at sufficiently high magnetic Reynolds numbers are cylindrical helical jets (Ponomarenko  dynamo \cite{Ponomarenko}), two-dimensional spatially periodic arrays of helical jets (Roberts' scheme \cite{Roberts}),  cylindrical von K\'arm\'an \cite{VKS_num} and Taylor-Couette flows \cite{TCF_num}, and spherical Dudley-James flows \cite{Dudley}. The first three of these flows have been tested recently in experiments with liquid sodium, and successful observations of  dynamo action have been reported in Refs.~\cite{Gailitis,Stieglitz,Monchaux}. The flows in these experiments were turbulent, thus making it difficult to compare experimental data with predictions of laminar dynamo theory, although in the first two, the role of turbulence was small as the flows were strongly constrained. This is a common disadvantage of all liquid metal dynamo experiments: the extremely low magnetic Prandtl numbers (the ratio of kinetic viscosity $\nu$ to resistivity $\eta$ or, equivalently, the ratio of magnetic Reynolds to fluid Reynolds $Pm=\nu/\eta=Rm/Re \sim 10^{-5}$ for liquid sodium) requires very high fluid Reynolds numbers ($Re\sim10^6-10^7$) in order to achieve the magnetic Reynolds number sufficient for dynamo action ($Rm\sim10^1-10^2$) in liquid metals. As a result the relevant flow is always turbulent.            

The present paper is motivated by the Madison Plasma Couette Experiment (MPCX) \cite{Collins}, which is designed to study MHD phenomena driven by plasma flows. One of the novelties of this experiment is the ability to change the magnetic Prandtl number of the plasma by several orders of magnitude from $Pm<<1$ to $Pm>>1$. This flexibility makes it possible to investigate laminar dynamos by  choosing a regime with $Pm\sim1-10$ and $Re\sim10^2$. As  a result the direct comparison of experimental data with numerical simulations of laminar dynamo can be performed.   Such a comparison can also be used for the first time to test different MHD models  as well as the numerical codes which simulate them.

The goal of this study is to investigate possible dynamo action in the MPCX using the extended MHD code NIMROD \cite{Sovinec}, which can accurately model plasma dynamics in the specific geometry for realistic experimental conditions. Among the features of NIMROD is the possibility to study effects beyond standard MHD, including the addition of the Hall term in Ohm's law. The effect of the Hall term on dynamo action has recently been studied in periodic box simulations  \cite{Mininni,Gomez}, and the Hall term will almost certainly play an important role in MPCX. The results of our simulations can also be used for the optimization of plasma parameters and as a guidance for the experimental operation.

In this paper we report the results of NIMROD simulations of laminar magnetic dynamo in the cylindrical von K\'arm\'an flow under conditions relevant to MPCX. Simulations are done for an isothermal compressible MHD plasma model with and without two-fluid effects (the Hall term). The structure of the paper is following. In section II we briefly review the MPCX experiment and describe the NIMROD plasma model. In section III the hydrodynamical properties of  von K\'arm\'an flow are studied.  In section IV the kinematic dynamo problem is considered for von K\'arm\'an flow in a cylinder, and the self-generation of the magnetic field is demonstrated for parameters achievable in the experiment. In section V the results of simulations of nonlinear dynamo saturation are presented, and the effect of the Hall term is studied. In section VI we conclude with a discussion of the effects that can influence the magnetic dynamo in such plasma flows. 

\section{NIMROD models of MPCX}

\begin{figure}[tbp]
\begin{center}
\includegraphics[scale=1]{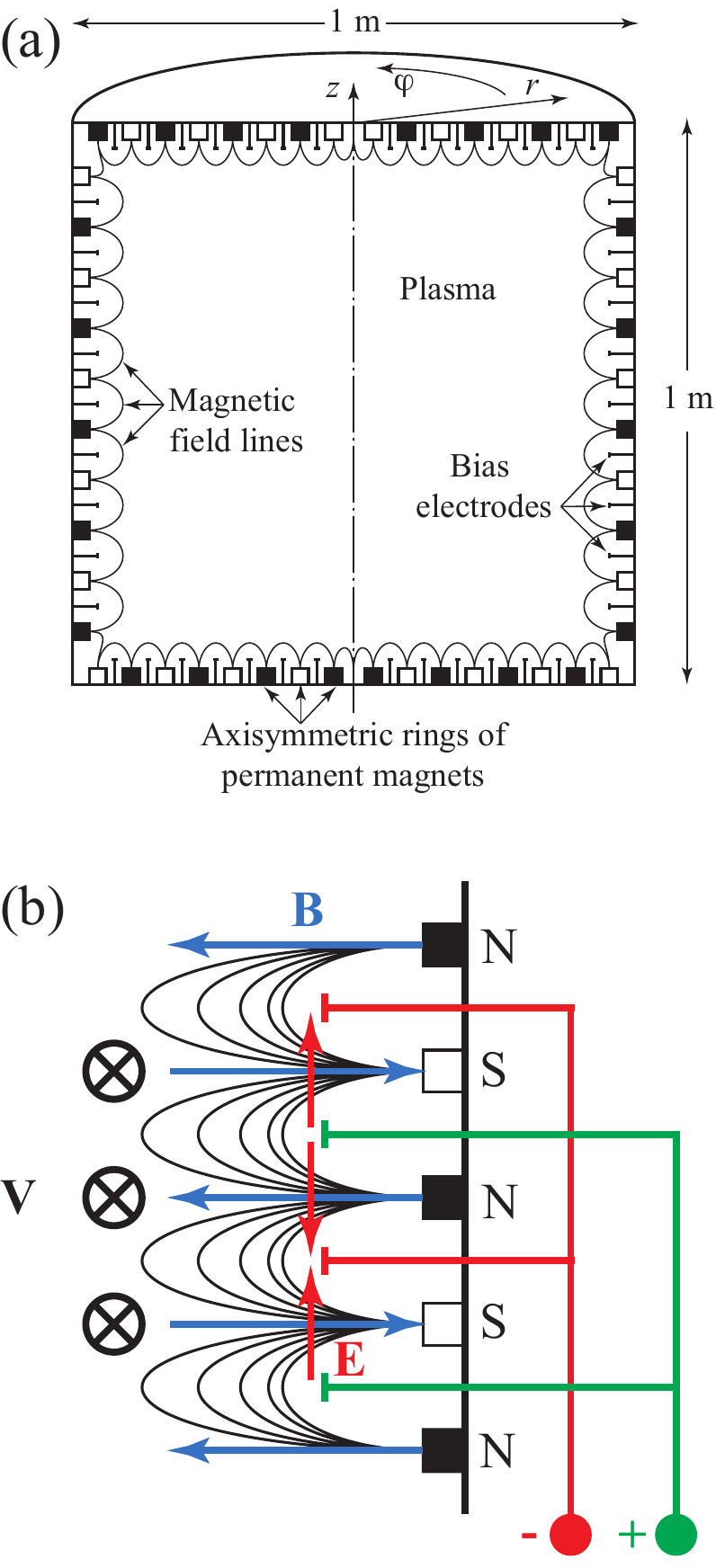}
\caption{Madison Plasma Couette Experiment (MPCX): (a) sketch; (b) partial vertical cross section. Rings of permanent magnets of alternating polarity line the inside of the cylinder with their poles oriented normally to the walls. Ring anodes and cathodes are placed between the magnets. The resulting $\mathbf{E}\times\mathbf{B}$ drift is in the azimuthal direction. By varying the potential between the anodes and cathodes the velocity forcing at the outer boundary can be customized.}\label{f1}
\end{center}
\end{figure}

MPCX is closely related to the spherical plasma experiment described in Ref.~\cite{Spence}, though here the geometry is cylindrical and the apparatus is somewhat smaller in size (1 m in diameter). In MPCX the plasma is confined by a multi-cusp magnetic field created by axisymmetric rings of permanent magnets of alternating polarity and localized at the boundary of the cylindrical chamber (Fig.~\ref{f1}). There are 10 magnetic rings at the cylindrical wall and 8 at both the top and bottom end-caps.  Ring anodes and cathodes positioned between the magnet rings can be biased with arbitrary potentials.  The resulting  $\mathbf{E}\times\mathbf{B}$ drift of plasma is in the  azimuthal direction and can be an arbitrary axisymmetric function at the boundary of the vessel. This arrangement allows  arbitrary shear flows to be imposed in the MPCX experiment.

The use of a plasma gives experimentalists great flexibility in choosing the regimes of operation. By varying the plasma density (gas flow rate), ion mass (H, He, Ne, Ar), electron temperature (heating power), and flow speed (bias potentials of electrodes), a wide range of parameters can be achieved in experiment (Table \ref{t1}). Such flexibility is advantageous over the liquid metal dynamo experiments, where $Pm$ is fixed and small. For description of the plasma parameters in our simulations we introduce several standard dimensionless numbers:
\begin{description}

\item Magnetic Prandtl:
\begin{equation}\label{Pm}
Pm\equiv\frac{\nu}{\eta}=46\,\frac{T_e^{3/2}[\textrm{eV}]\,T_i^{5/2}[\textrm{eV}]}{n_0[10^{18}\,\textrm{m}^{-3}]\,\mu_i^{1/2}\lambda^2},
\end{equation}

\item Fluid Reynolds:
\begin{equation}\label{Re}
Re\equiv\frac{R_0V_0}{\nu}=0.52\frac{R_0[\textrm{m}]\,V_0[\textrm{km/s}]\,n_0[10^{18}\,\textrm{m}^{-3}]\,\mu_i^{1/2}\lambda}{T_i^{5/2}[\textrm{eV}]},
\end{equation}

\item Magnetic Reynolds:
\begin{equation}\label{Rm}
Rm\equiv\frac{R_0V_0}{\eta}=24\,\frac{R_0[\textrm{m}]\,V_0[\textrm{km/s}]\,T_e^{3/2}[\textrm{eV}]}{\lambda},
\end{equation}

\item Mach:
\begin{equation}\label{M}
M\equiv\frac{V_0}{C_s}=0.10\frac{V_0[\textrm{km/s}]\,\mu_i^{1/2}}{\gamma^{1/2}\,T_e^{1/2}[\textrm{eV}]},
\end{equation}

\item Hall:
\begin{equation}\label{eps}
\varepsilon\equiv\frac{c}{R_0\omega_{pi}}=0.23\frac{\mu_i^{1/2}}{R_0[\textrm{m}]\,n_0^{1/2}[10^{18}\,\textrm{m}^{-3}]},
\end{equation}

\end{description}
where $\nu$ is the plasma kinematic viscosity,  $\eta$ is the magnetic diffusivity, $\lambda$ is the Coulomb logarithm (typically $\lambda\approx10-20$), $C_s$ is the ion sound speed, $\gamma$ is the adiabatic index, $c$ is the speed of light, $\omega_{pi}$ is the ion plasma frequency and other parameters are defined in Table \ref{t1}. Formulas (\ref{Pm})-(\ref{Rm}) for numerical estimates of $Pm$, $Re$ and $Rm$ are derived from Braginskii equations for a plasma with singly charged ions in a weak magnetic field  \cite{Brag}; the weak-field approximation is reasonable for the bulk of MPCX because the high-multipole cusp field is concentrated mostly near the wall and quickly falls off away from it. The typical values of these non-dimensional numbers are listed in Table \ref{t1}. For convenience we also give the ``inverse" mapping formulary:  
\begin{description}

\item Peak driving velocity, km/s:
\begin{equation}
V_0=2.54\,\frac{\lambda^{1/4}\gamma^{3/8}Rm^{1/4}M^{3/4}}{\mu_i^{3/8}R_0^{\,1/4}[\textrm{m}]},
\end{equation}

\item Average number density, $10^{18}$ m$^{-3}$:
\begin{equation}
n_0=0.053\,\frac{\mu_i}{\varepsilon^2R_0^{\,2}[\textrm{m}]},
\end{equation}

\item Electron temperature, eV:
\begin{equation}
T_e=0.065\,\frac{\lambda^{1/2}\mu_i^{1/4}Rm^{1/2}}{\gamma^{1/4}M^{1/2}R_0^{\,1/2}[\textrm{m}]},
\end{equation}

\item Ion temperature, eV:
\begin{equation}
T_i=0.35\,\frac{\lambda^{1/2}\mu_i^{9/20}\gamma^{3/20}Rm^{1/10}M^{3/10}}{\varepsilon^{4/5}Re^{2/5}R_0^{\,1/2}[\textrm{m}]}.
\end{equation}

\end{description}

\begin{table}[tbp]
\caption{Parameters of MPCX}
\begin{center}
\begin{tabular}{lccc}
\hline
\hline
Quantity & Symbol & Value & Unit \\
\hline 
Radius of cylinder & $R_0$ & 0.5 & m \\
Height of cylinder & $H$ & 1 & m \\
Peak driving velocity  & $V_0$ & 0-20 & km/s\\
Average number density &   $n_0$  &  $10^{17}-10^{19}$ & m$^{-3}$\\
Electron temperature & $T_e$ & 2-10 & eV \\
Ion temperature & $T_i$ &  $0.5-4$ & eV \\
Ion species &  & H, He, Ne, Ar &\\
Ion mass & $\mu_i$ & 1, 4, 20, 40 & amu\\
Ion charge & $Z$ & 1 & e\\
\hline
Magnetic Prandtl & $Pm$ & $1\times10^{-3}-5\times10^2$ & \\
Fluid Reynolds  & $Re$ & $0-4\times10^4$ & \\
Magnetic Reynolds & $Rm$ & $0-1\times10^3$ & \\
Mach & $M$ & $0-4$ & \\
Hall  & $\varepsilon$ & $0.15-1.8$ & \\
\hline
\end{tabular}
\end{center}
\label{t1}
\end{table}

Results presented in this paper are obtained using the extended MHD code NIMROD \cite{Sovinec}. As a simulation framework, we choose the isothermal Hall MHD approach.  This is one of the simplest NIMROD models allowing for two-fluid effects and compressibility, and this appears to be a good approximation for the plasma under experimental conditions. The equations of this model in non-dimensional form are:
\begin{eqnarray}
\label{model1}
\frac{\pa n}{\pa\tau}&=&-\nabla\cdot(n\v),\\
\label{model2}
n\frac{\pa \v}{\pa\tau}&=&-n(\v\cdot\nabla)\v - \frac{1}{M^2}\nabla n+(\nabla\times\b)\times\b + \frac{1}{Re}\bigg(\nabla^2\v+\frac{1}{3}\nabla(\nabla\cdot\v)\bigg),\\
\label{model3}
\frac{\pa\b}{\pa\tau}&=&\nabla\times\bigg(\v\times\b-\frac{\varepsilon}{n}(\nabla\times\b)\times\b\bigg)+\frac{1}{Rm}\nabla^2\b.
\end{eqnarray}
In these equations $\tau$, $n$, $\v$ and $\b$ stand for normalized time, number density, velocity and magnetic field, respectively: 
$$
\tau=\frac{V_0}{R_0}\,t,~~~n=\frac{\rho}{n_0m_i},~~~\v=\frac{\V}{V_0},~~~\b=\frac{\B}{V_0\sqrt{4\pi n_0m_i}},
$$
where $\rho$ is the mass density, $m_i$ is the ion mass. The unit of length is the cylinder radius  $R_0$, while $V_0$ is the peak velocity of the driven von K\'arm\'an flow. An important difference of this system of equations (\ref{model1})-(\ref{model3}) from a standard single-fluid MHD model is the inclusion of the Hall term in the magnetic induction equation (\ref{model3}), which takes into account two-fluid effects.  The magnitude of this term is characterized by the non-dimensional Hall number $\varepsilon$. We will consider simulations where the Hall term is significant ($0<\varepsilon<1$ ) and others where it is absent ($\varepsilon = 0$). The simulations are performed in a non-rotating cylindrical coordinate system  $(r,~\varphi,~z)$, with the plasma occupying the region $(0<r<1,~-1<z<1)$. 

\begin{figure}[tbp]
\begin{center}
\includegraphics[scale=0.9]{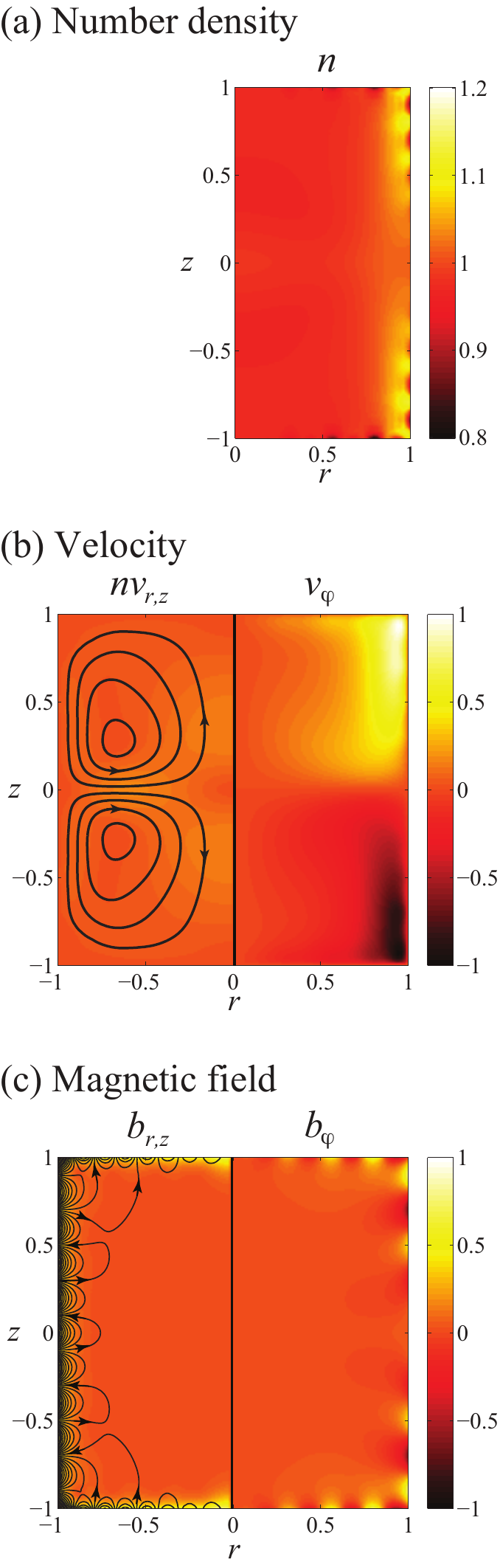}
\caption{Structure of axisymmetric equilibrium von K\'arm\'an flow driven by electro-magnetic system at the boundary for Mach number $M=1$, fluid Reynolds number $Re=200$ and magnetic Reynolds number $Rm=20$: (a) number density; (b) velocity; (c) magnetic field. Cross-sections in $r-z$ plane are given. Left panels of (b) and (c) show stream lines of poloidal parts ($r$- and $z$-components) of flux $n\v$ and magnetic field $\b$, respectively, superimposed on absolute values of these parts depicted in colors. Right panels of (b) and (c) show azimuthal components of corresponding fields.}\label{f2}
\end{center}
\end{figure}

The Hall MHD equations (\ref{model1})-(\ref{model3}) also require the specification of boundary conditions. Two different sets of boundary conditions are used in the simulations. Set I is used only to demonstrate the possibility of stirring the plasma with the applied multi-cusp magnetic field and an appropriately modulated tangential electric field at the boundary. In set I no-slip, stationary, rigid walls are assumed, so all components of the velocity vanish at the boundary,
\begin{equation}\label{BCv1}
\v|_\Gamma=0.
\end{equation}   
For the magnetic and electric fields in set I, we assume perfectly conducting walls, implying that the time-varying normal component of the magnetic field and the time-varying tangential component of the electric field are zero at the boundary:
\begin{equation}\label{BCb1}
\tilde{b}_n|_\Gamma=0,~~~\tilde{E}_t|_\Gamma=0,
\end{equation}   
where the normalized electric field is 
$$
\E=-\v\times\b+\frac{\varepsilon}{n}(\nabla\times\b)\times\b+\frac{1}{Rm}\nabla\times\b.
$$ 
Note that the externally applied time-independent multi-cusp magnetic field $\b_0$ and boundary electric field $\E_0$ do not satisfy conditions (\ref{BCb1}), and thus provide $\E_0 \times \b_0$ stirring at the boundary. Using different modulations of the tangential boundary electric field we have successfully simulated several types of flows. Fig.~\ref{f2} shows the results for the so-called von K\'arm\'an flow, in which plasma is driven in opposite azimuthal directions near the top and bottom end-caps.  Having demonstrated that $\E_0 \times \b_0$ stirring will successfully drive a von K\'arm\'an flow under realistic experimental conditions, we now turn to a simpler set of boundary conditions.

All results reported in the later sections are obtained with the boundary conditions of set II. In set II we ignore the applied multi-cusp magnetic field and the tangential electric field. Instead we assume that the driving of the plasma is due to differentially rotating walls. This assumption greatly simplifies the model and allows us to focus on the physics of the dynamo action and not on details of the plasma driving. The boundary conditions for the full electric and magnetic fields are:
\begin{equation}\label{BCb2}
b_n|_\Gamma=0,~~~E_t|_\Gamma=0
\end{equation}      
(perfectly conducting walls), and the velocity conditions are:
\begin{eqnarray}\label{BCv2}
\v|_{r=1}&=&z\,\e_\varphi,\\
\v|_{z=-1}&=&-r\,\e_\varphi,\nonumber\\
\v|_{z=1}&=&r\,\e_\varphi \nonumber
\end{eqnarray}
(no-slip differentially rotating rigid walls). These velocity boundary conditions correspond approximately to von K\'arm\'an flow: the top and bottom end-caps are counter-rotating with the same angular velocity, and the side wall has a linear dependence of angular velocity on $z$ to match the rotation of top and bottom end-caps. This flow is the primary object of our dynamo study.

We briefly remark on the spatial and temporal resolution used in these simulations. For spatial discretization NIMROD employs a high order finite element method in $r$- and $z$-directions and a pseudospectral method in periodic $\varphi$-direction with dependence $e^{im\varphi}$ for each Fourier harmonic (integer $m$ represents the azimuthal mode number). The basis functions of the finite elements are polynomials. In all of the simulations presented here we have used a uniform meshing of the $r-z$ plane with $8\times16$ finite elements each of polynomial degree 3, and 11 Fourier harmonics in the azimuthal $\varphi$-direction. This spatial resolution appears to be sufficient for the laminar flows under consideration. To verify the simulation results obtained at this resolution, we have repeated a nonlinear MHD run introduced in section V using mesh with $16\times32$ finite elements in $r-z$ plane and 11 azimuthal modes. The time dynamics of the flow and the magnetic field  were fully reproduced, which confirms that a converged solution is obtained already at the coarser resolution.  The solutions are time-evolved using a semi-implicit staggered leap-frogging algorithm, which is fully detailed in Refs. \cite{Sovinec, Sovinec_2010}. We should emphasize here that the algorithm employed in NIMROD can be made numerically stable for arbitrarily large time-steps in both single-fluid and Hall MHD by choosing the appropriate coefficients of the semi-implicit operators \cite{Sovinec_2010}. However, in order to accurately model the  temporal behavior of the system with significant flows, in the present simulations we have used an adaptive time-step based on the explicit Courant-Friedrichs-Lewy (CFL) condition for advection.  Even in the dynamo simulations, we find that advection by the axisymmetric velocity field dominates the CFL criteria, rather than either the Alfv\'en or whistler waves associated with the relatively weak magnetic fields.

\section{Hydrodynamical equilibrium and stability}

\begin{figure}[tbp]
\begin{center}
\includegraphics[scale=0.9]{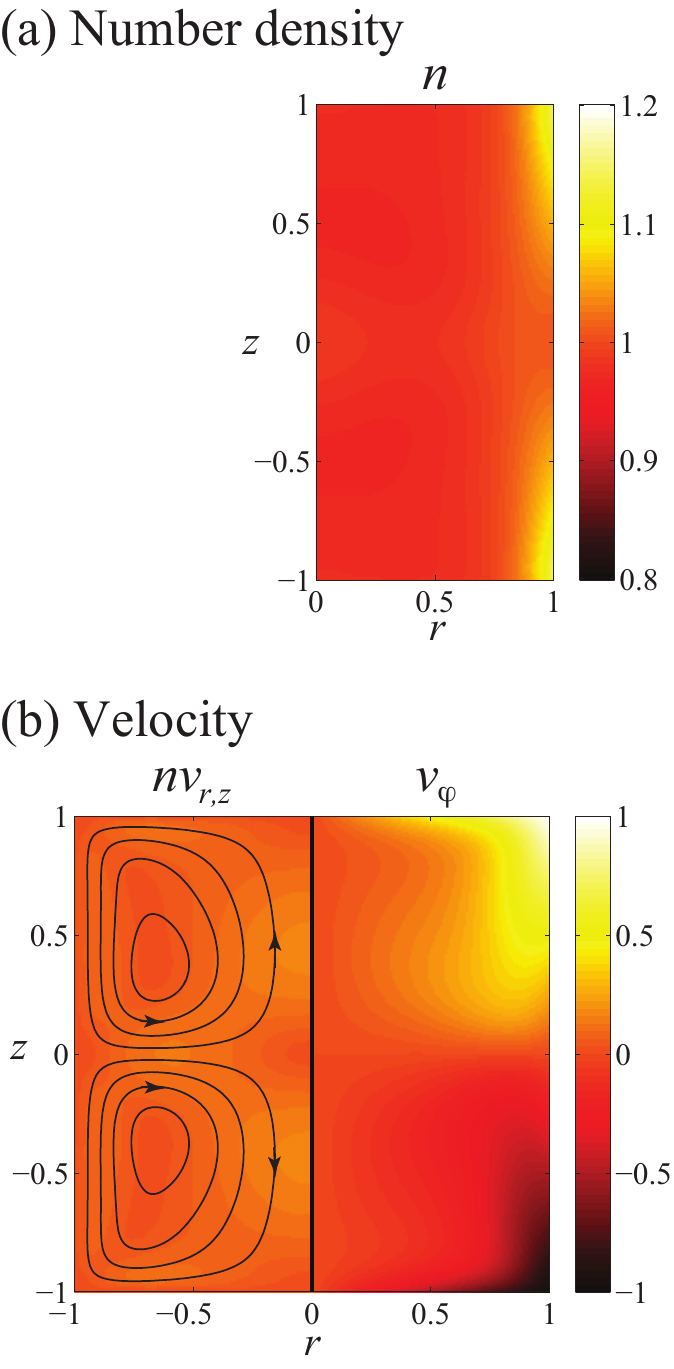}
\end{center}
\caption{Structure of axisymmetric equilibrium von K\'arm\'an flow driven by differentially rotating walls for Mach number $M=1$ and fluid Reynolds number $Re=200$: (a) number density; (b) velocity. The same elements as in Fig.~\ref{f2} are shown.}\label{f3}
\end{figure}

In this section we consider the basic hydrodynamical (no magnetic field, $\b=0$) properties of von K\'arm\'an flow in cylinder, in which the plasma is stirred at the edge via the boundary conditions of set II (\ref{BCv2}). We start with axisymmetric case, assuming that physical quantities do not depend on $\varphi$, i.e., $\pa/\pa\varphi=0$. Axisymmetric equilibrium flow structure for fluid Reynolds number $Re=200$ and Mach number $M=1$  is shown in Fig.~\ref{f3}. The velocity components are either symmetric (radial $v_r$) or antisymmetric (azimuthal $v_\varphi$ and axial $v_z$) with respect to equatorial plane $z=0$, and the azimuthal velocity has maxima at the corners of the cylinder. The flow develops two cells of poloidal (in the $r-z$ plane) circulation with inward direction at the equatorial plane. Such flow pattern leads to the stratification of density, which builds up near the corners of the cylinder.

\begin{figure}[tbp]
\begin{center}
\includegraphics[scale=1]{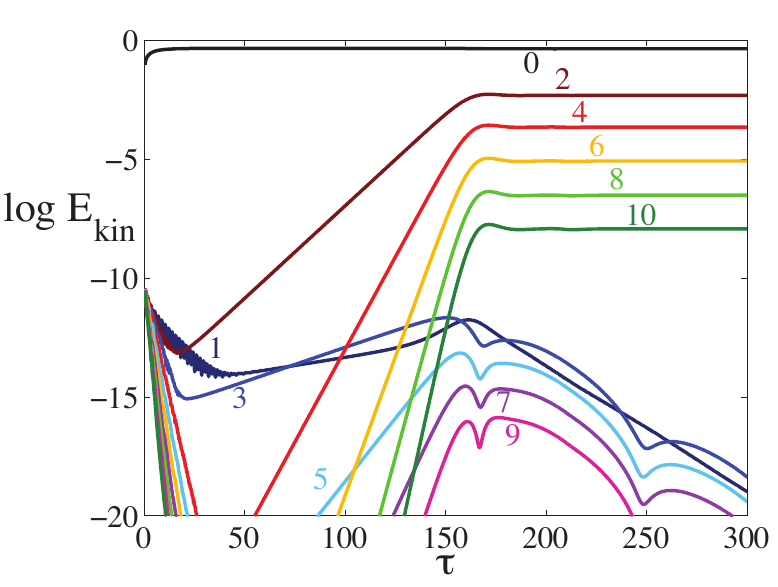}
\end{center}
\caption{Time dynamics of kinetic energy of different azimuthal harmonics in purely hydrodynamical ($\b=0$)  von K\'arm\'an flow for Mach number $M=1$ and fluid Reynolds number $Re=200$. Corresponding azimuthal mode numbers $m$ are shown.}\label{f4}
\end{figure}

The axisymmetric equilibrium von K\'arm\'an flow becomes unstable with respect to non-axisymmetric perturbations when the fluid Reynolds exceeds a critical value (Fig.~\ref{f4}). This instability is the Kelvin-Helmholtz type instability, occurring due to the presence of unstable velocity shear in the fluid. For Mach number $M=1$ the critical value of Reynolds number is about $Re\approx160$. As shown in Fig.~\ref{f4}, within approximately one viscous time the unstable modes grow and saturate at a new equilibrium state, which consists of the axisymmetric part and relatively small non-axisymmetric distortions with even azimuthal mode numbers, $m=2,~4,~6,\ldots$. Such transition to non-axisymmetric equilibrium is crucial for the dynamos considered here.

\begin{figure}[btp]
\begin{center}
\includegraphics[scale=1]{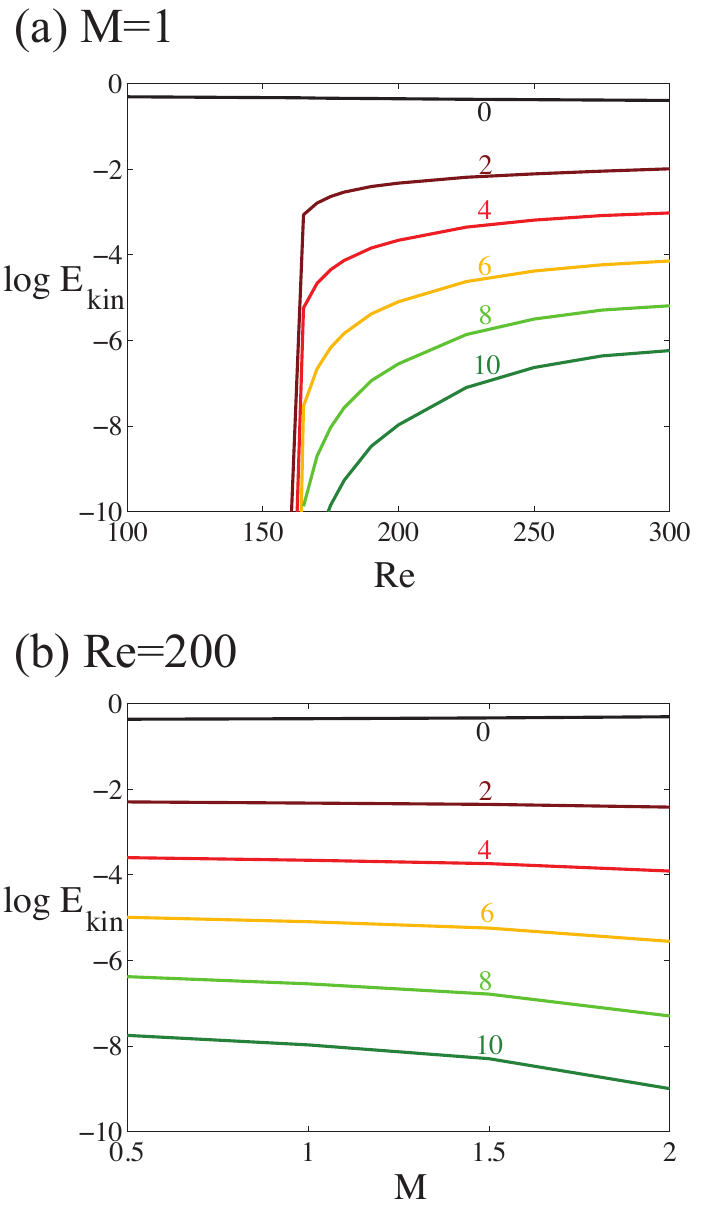}
\end{center}
\caption{Kinetic energy of different azimuthal harmonics in hydrodynamically stable von K\'arm\'an flow as a function of (a) fluid Reynolds number $Re$ for Mach number $M=1$; (b) Mach number $M$ for fluid Reynolds number $Re=200$. Corresponding azimuthal mode numbers $m$ are shown.}\label{f5}
\end{figure}

The structure of the equilibrium von K\'arm\'an flow, in particular the amplitude of the non-axisymmetric distortions, depends on Reynolds and Mach numbers (Fig.~\ref{f5}). As we discuss in section IV, such dependence affects the critical value of magnetic Reynolds number above which the dynamo is excited.      

\section{Kinematic dynamo}

\begin{figure}[tbp]
\begin{center}
\includegraphics[scale=1]{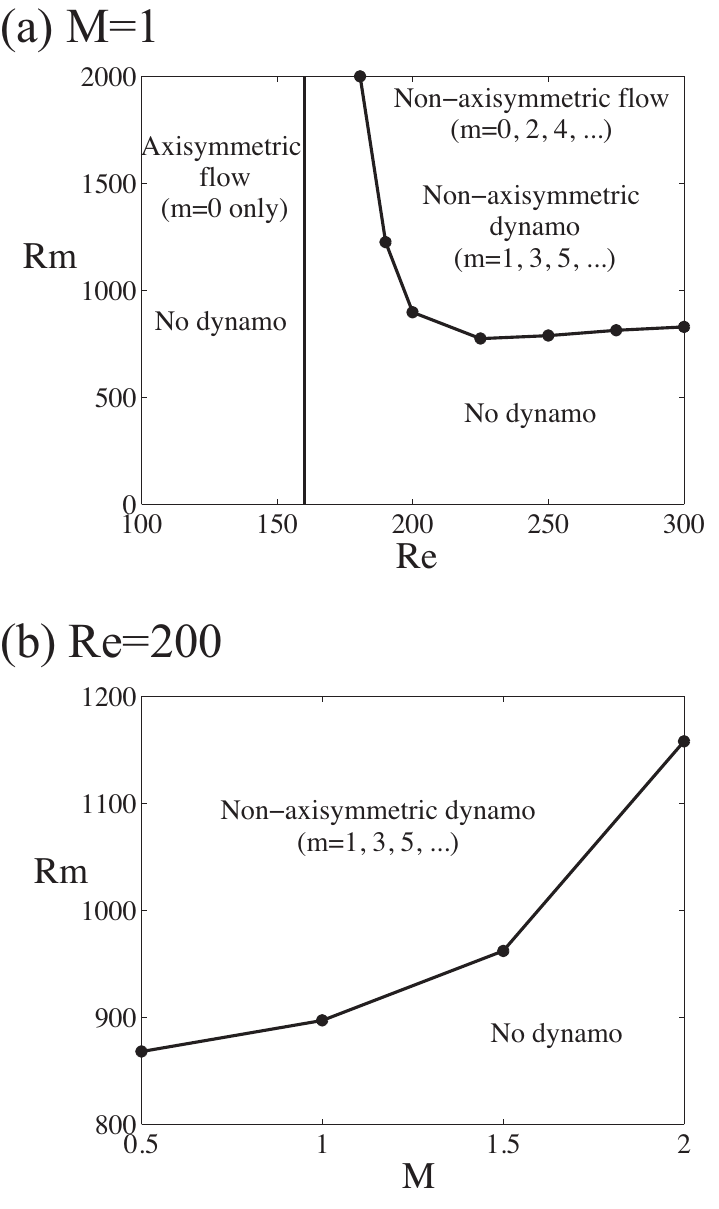}
\end{center}
\caption{Critical magnetic Reynolds number $Rm$ as a function of (a) fluid Reynolds number $Re$ for Mach number $M=1$; (b) Mach number $M$ for fluid Reynolds number $Re=200$. Vertical line in (a) separates the regions of axisymmetric and non-axisymmetric equilibrium von K\'arm\'an flows.}\label{f6}
\end{figure}

Our first step in this dynamo study is to solve the kinematic dynamo problem: determining the possibility of self-generation of magnetic field for a given flow structure. For fixed fluid Reynolds and Mach numbers we solve the stationary time independent ($\pa/\pa\tau=0$) continuity (\ref{model1}) and Navier-Stokes equations (\ref{model2}) and find the steady-state hydrodynamic equilibrium velocity $\v_\mathrm{eq}$ (which includes possible non-axisymmetric distortions). Using that velocity, we solve induction equation (\ref{model3}) as an eigenvalue problem for magnetic field with different magnetic Reynolds numbers 
\begin{equation}
\label{kin}
\gamma\b=\nabla\times(\v_\mathrm{eq}\times\b)+\frac{1}{Rm}\nabla^2\b,
\end{equation}
where $\gamma$ is an eigenvalue. This allows us to determine the critical magnetic Reynolds above which the dynamo excitation is possible. The results are presented in Fig.~\ref{f6}. The dependence of the critical magnetic Reynolds on the fluid Reynolds for Mach number $M=1$ is shown in Fig.~\ref{f6}(a). The vertical line at $Re\approx160$ separates the regions of axisymmetric ($Re<160$) and non-axisymmetric ($Re>160$) von K\'arm\'an flows. Our simulations show that  the kinematic dynamos are not possible in axisymmetric flows. In non-axisymmetric flow at sufficiently high magnetic Reynolds number $Rm$, the dynamo appears as a growing magnetic field with odd azimuthal harmonics, $m=1,~3,~5,\ldots$.  

In our study of the plasma dynamo we use a compressible fluid model. Compressibility is related to the Mach number (\ref{M}) -- the ratio of the peak driving velocity to the sound speed. In general, the higher Mach number the more compressible the fluid, i.e., the more stratified its density. An increase in the Mach number leads to changes in the equilibrium velocity field $\v_\mathrm{eq}$ and, in particular, decreases the energy in the non-axisymmetric components of the flow (Fig.~\ref{f5}(b)). This, in turn, affects the kinematic dynamo problem (\ref{kin}) by increasing the value of the critical $Rm$ (Fig.~\ref{f6}(b)). 

From a comparison of Figs.~\ref{f5} and \ref{f6} we can conclude that the dynamo action in the case under consideration is related to the non-axisymmetric part of the von K\'arm\'an flow: the stronger the non-axisymmetric distortions, the lower the threshold value of magnetic Reynolds number for the onset of the dynamo.  

\section{Non-linear saturation of dynamo field}

Next we analyze the state of  the fully saturated dynamo generated magnetic field and its back reaction on the flow. In this section we report the results of fully non-linear simulations of system (\ref{model1})-(\ref{model3}). The non-dimensional parameters are chosen to be $Re=200$, $Rm=1000$, $M=1$ and $\varepsilon=0.05-1.0$ (for Hall MHD runs); simulations with these parameters achieve fully saturated laminar states for the dynamo fields and the fluid flows.

\begin{figure*}[btp]
\begin{center}
\includegraphics[scale=1]{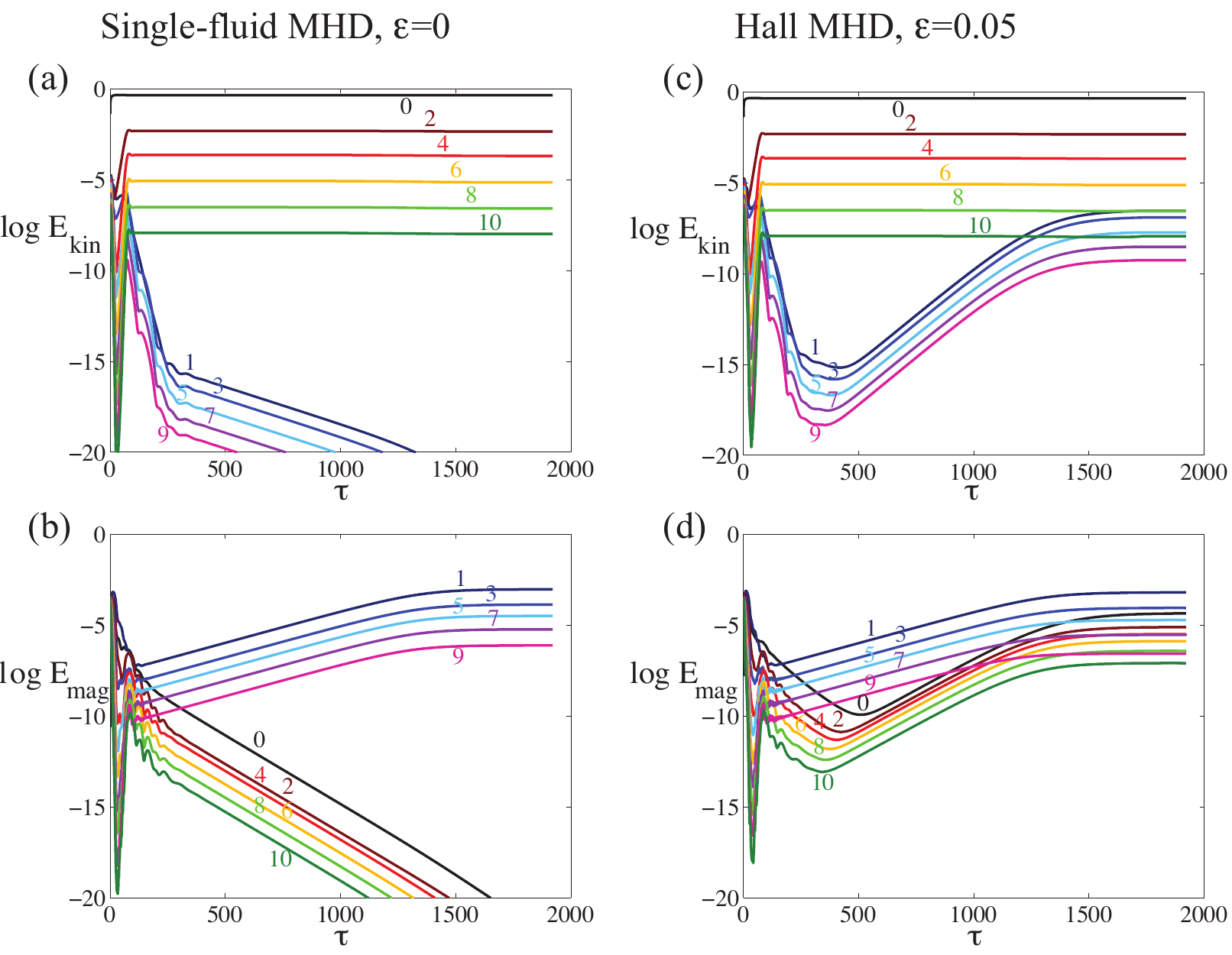}
\end{center}
\caption{Time dynamics of kinetic $E_\mathrm{kin}$ and magnetic $E_\mathrm{mag}$ energies of different azimuthal modes in von K\'arm\'an flow for Mach number $M=1$, fluid Reynolds $Re=200$, and magnetic Reynolds $Rm=1000$, with azimuthal mode numbers $m$ labelled. (a,b) Single-fluid MHD case ($\varepsilon=0$).  Flows are of even $m$ modes while fields are odd in $m$. (c,d) Hall MHD case ($\varepsilon=0.05$).  Initial behavior is similar to the single-fluid MHD case, but as the fields become strong ($\tau \approx 500$), the Hall effect becomes important.  The final equilibrium includes both odd and even $m$.}\label{f7}
\end{figure*}

\begin{figure}[tbp]
\begin{center}
\includegraphics[scale=1]{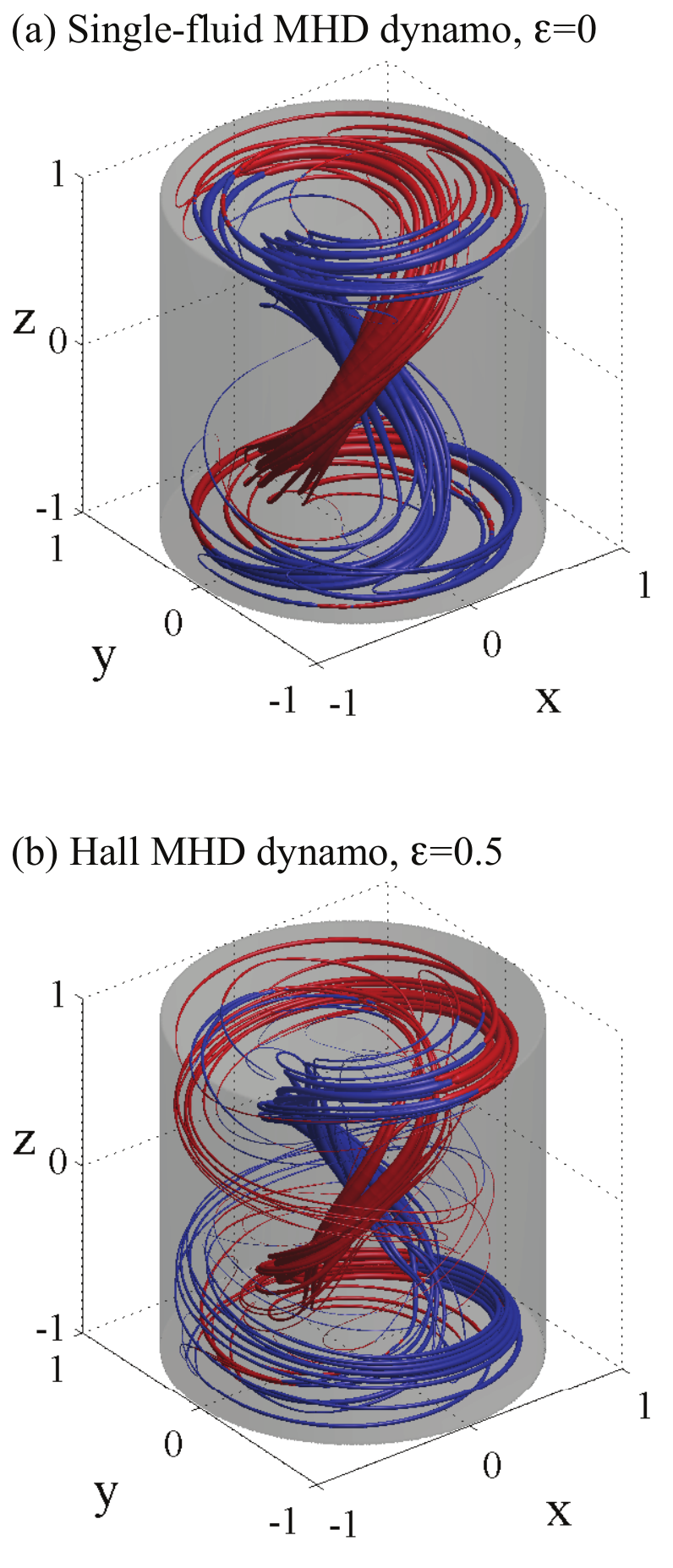}
\end{center}
\caption{Magnetic field lines of saturated dynamos: (a) single-fluid MHD case ($\varepsilon=0$); (b) Hall MHD case  ($\varepsilon=0.5$). Thickness of the line is proportional to the magnitude of the field, while red (blue) color corresponds to upward  (downward) direction of the field.}\label{f8}
\end{figure}

Figs.~\ref{f7}(a,b) demonstrate the time dynamics of the kinetic and magnetic energies for a single-fluid MHD case ($\varepsilon=0$). After the initial transient phase ($\tau\approx100$), the flow becomes stationary. It is primarily axisymmetric, with non-axisymmetric distortion consisting of even azimuthal harmonics ($m=2,~4,~6,\ldots$) and containing only about $1\%$ of the total kinetic energy. In such a flow the dynamo is excited: magnetic field of odd harmonics ($m=1,~3,~5,\ldots$) grows exponentially in time until it saturates at $\tau\approx1500$. In some sense, the saturated magnetic field is in  equipartition with the non-axisymmetric part of the flow. Here $E_\mathrm{mag}$ is about 0.3$\%$ of the total energy of the flow $E_\mathrm{kin}$. Due to the lack of an axisymmetric magnetic field ($m=0$) and the small amplitude of the non-axisymmetric fields, the back reaction of the dynamo magnetic field on the axisymmetric flow is very weak and the imposed von K\'arm\'an flow is essentially unmodified. The structure of the saturated magnetic field, with the $m=1$ azimuthal harmonic visibly dominating the overall structure, is shown in Fig.~\ref{f8}(a).

In the Hall MHD case  ($\varepsilon=0.05$) the dynamics of the system is different (Figs.~\ref{f7}(c,d)). After the non-axisymmetric flow with even modes develops (at $\tau\approx100$) the odd harmonics of the magnetic field start to grow, as in single-fluid MHD case. Now however, when the odd harmonics of the magnetic field become large enough (at $\tau\approx500$), the Hall effect becomes significant and other harmonics of the flow and the magnetic field grow, breaking the initial even-odd symmetry. The saturated state of the flow and the magnetic field contains all harmonics, though the dominating parts are similar to MHD case. The structure of the saturated Hall dynamo field for $\varepsilon=0.5$ is shown in Fig.~\ref{f8}(b).

\begin{figure}[tbp]
\begin{center}
\includegraphics[scale=1]{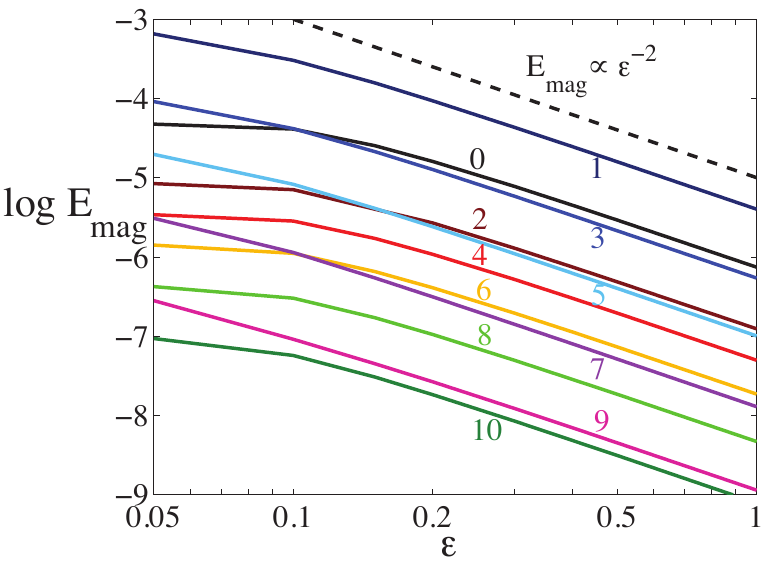}
\end{center}
\caption{Magnetic energy of different azimuthal harmonics in saturated Hall MHD dynamo as a function of the Hall number $\varepsilon$.  Azimuthal mode numbers $m$ are shown. Dashed line corresponds to scaling $E_\mathrm{mag}\propto\varepsilon^{-2}$.}\label{f9}
\end{figure}

Fig.~\ref{f9} shows the dependence of the magnetic energy for different azimuthal modes in the saturated state on the Hall number $\varepsilon$. It is worth noting that the presence of the Hall effect not only breaks the even-odd symmetry of the system, but also reduces the energy of the saturated dynamo field. For $\varepsilon\gtrsim0.2$ the energy of the saturated dynamo field scales as $E_\mathrm{mag}\propto\varepsilon^{-2}$ ($b\propto\varepsilon^{-1}$ for the field amplitude). This is a direct consequence of the magnetic induction equation (\ref{model3}), which in a saturated state ($\pa/\pa\tau=0$) now reads:
\begin{equation}
\label{steady}
\nabla\times\bigg(\v\times\b-\frac{\varepsilon}{n}(\nabla\times\b)\times\b\bigg)+\frac{1}{Rm}\nabla^2\b=0.
\end{equation}
Indeed, if the magnetic Reynolds number $Rm$ and functions $\v$ and $n$ are independent of $\varepsilon$, then the solution for the magnetic field $\b$ can be written as
\begin{equation}
\label{steady2}
\b=\frac{\bh}{\varepsilon},
\end{equation}
where $\bh$ is a vector-function independent of $\varepsilon$ satisfying equation 
\begin{equation}
\label{steady3}
\nabla\times\bigg(\v\times\bh-\frac{1}{n}(\nabla\times\bh)\times\bh\bigg)+\frac{1}{Rm}\nabla^2\bh=0
\end{equation}
and boundary conditions (\ref{BCb2}).  In the case under consideration it is clear that the scaling used in (\ref{steady2}) is valid only asymptotically for large Hall numbers $\varepsilon$ when the magnetic field $\b$ is small and its influence on the flow is negligible.  Under these assumptions, the plasma velocity $\v$ and density $n$ do not depend on $\varepsilon$ and correspond to the purely hydrodynamical equilibrium that is determined by the fluid Reynolds $Re$ and Mach $M$ numbers (as described in section III). Thus, for relatively large Hall parameters ($\varepsilon\gtrsim0.2$)  the saturation of the dynamo takes place due to the back reaction of the Hall term in induction equation (\ref{model3}) long before the amplitude of the magnetic field is large enough to change the flow significantly. This is in contrast to the single-fluid MHD dynamo (and the Hall dynamo with $\varepsilon<<0.1$),  where the saturated state is achieved due to modification of the flow profile by the growing dynamo field before the Hall effect plays a considerable role. Scaling (\ref{steady2}) suggests that the Hall effect is unfavorable for the dynamo.     

We note Refs.~\cite{Bayliss,Reuter,Giss}, where results of simulations for a single-fluid MHD dynamo in a sphere with counter-rotating incompressible flows are reported. Despite the difference in geometry, our MHD results are in qualitative agreement with the cases of laminar dynamos from Refs.~\cite{Bayliss,Reuter}, which have dominant $m=0$ and $m=2$ harmonics in kinetic energy and $m=1$ harmonic in magnetic energy during the saturation phase. However, in contrast to results of Ref.~\cite{Giss}, we do not observe the generation of an axial magnetic dipole with $m=0$ and quasi-periodic oscillations.

\section{Discussion}

We have performed numerical simulations of a plasma dynamo for cylindrical von K\'arm\'an flow using the extended MHD code NIMROD. These NIMROD simulations provide numerical support for the Madison Plasma Couette Experiment (MPCX). We have demonstrated that sustained dynamo action and self-generation of magnetic field can be attained for parameters that are achievable in the experiment. Our results show that the critical magnetic Reynolds number required for dynamo excitation strongly depends on the plasma compressibility: the more compressible the fluid, the higher the critical $Rm$ (Fig.~\ref{f6}). Inclusion of two-fluid effects into the  model (in the form of the Hall term in the induction equation) does not influence the critical $Rm$, but does change the structure of the saturated flow and the dynamo field (Fig.~\ref{f7}). The effect of the Hall term on dynamo field is negative: the energy of the saturated magnetic field scales as $1/\varepsilon^2$ when the Hall number is $\varepsilon\gtrsim0.2$ (Fig.~\ref{f9}).     

The simulations show that the presence of non-axisymmetric distortions in the flow plays a decisive role in dynamo excitation. Such distortions with even azimuthal mode numbers ($m=2,~4,~6,\ldots$) appear in the flow only for modest values of fluid Reynolds number $Re>160$ when the axisymmetric shear flow becomes hydrodynamically unstable. Therefore, in order to observe  the dynamo effect in the MPCX, the plasma has to be driven above a critical hydrodynamical threshold. For a helium plasma with number density $n_0=10^{18}$ m$^{-3}$, electron temperature $T_e=16$ eV and ion temperature $T_i=0.9$ eV, the peak driving velocity should be $V_0=20$ km/s. 

Another issue is related to the detection of the laminar dynamo field in the experiment. Since the saturated magnetic energy of the dynamo field is only a small fraction  of the total kinetic energy, the resulting dynamo field is relatively weak. For the helium plasma considered above, the volume-averaged saturated dynamo magnetic field is $B_0\approx0.1$ Gauss. Such a small field is still detectable even on the background of the much stronger multi-cusp field from the rings of magnets (about $10^3$ Gauss near the walls) due to the different azimuthal symmetries of these two fields.  

Lastly, we remark on the model used in our simulations. The single-fluid MHD model ($\varepsilon =0$) is adequate for predicting critical magnetic Reynolds numbers and the thresholds for sustained dynamo action.  The fully non-linear saturated dynamo state differs however when the Hall effect is included.  It appears that the isothermal Hall MHD model ($\varepsilon > 0$) is a good ``rough" approximation for the plasma under experimental conditions, but this model also does not capture the full details of the plasma dynamics. Other effects such as thermal conductivity, electron pressure, and anisotropic viscosity can all play important roles in a real plasma experiment. These effects will be the subject of future studies.     

\acknowledgements

The authors  wish to thank Dr. C. Sovinec for valuable help and discussions related to \mbox{NIMROD}. The work is supported by the National Science Foundation.

\end{document}